\documentclass[article,12pt]{extarticle}
\usepackage{mathrsfs} 
\usepackage{amsmath} 
\usepackage{amssymb} 
\usepackage{mathtools}
\usepackage{hyperref} 
\usepackage{float}
\usepackage{endnotes} 
\usepackage[utf8]{inputenc}
\usepackage{graphicx}
\usepackage{setspace} 
\usepackage[margin=1in]{geometry} 

\newtheorem{theorem}{Theorem}[section]
\newtheorem{corollary}{Corollary}[theorem]

\newtheorem{proposition}{Proposition}[section] 
\newtheorem{definition}{Definition}[section] 
\newtheorem{example}{Example}[section]

\usepackage[style=apa, backend=biber]{biblatex} 
\begin{document}
\thispagestyle{empty} 

\begin{center}
    \vspace*{2cm}
    \huge\textbf{Mathematics of Differential Machine Learning in Derivative Pricing and Hedging}\par
    \vspace{1.5cm}
    \large\textbf{Running Title: Diff. ML in Derivative Pricing}\par
    \vspace{1cm}
    \large Pedro Duarte Gomes\par
    \vspace{0.5cm}
    \footnotesize Department of Mathematics, University of Copenhagen\par
    \vspace{0.3cm}
\end{center}

\newpage

\begin{abstract}

This article introduces the groundbreaking concept of the financial differential machine learning algorithm through a rigorous mathematical framework. Diverging from existing literature on financial machine learning, the work highlights the profound implications of theoretical assumptions within financial models on the construction of machine learning algorithms.

This endeavour is particularly timely as the finance landscape witnesses a surge in interest towards data-driven models for the valuation and hedging of derivative products. Notably, the predictive capabilities of neural networks have garnered substantial attention in both academic research and practical financial applications.

The approach offers a unified theoretical foundation that facilitates comprehensive comparisons, both at a theoretical level and in experimental outcomes. Importantly, this theoretical grounding lends substantial weight to the experimental results, affirming the differential machine learning method's optimality within the prevailing context.

By anchoring the insights in rigorous mathematics, the article bridges the gap between abstract financial concepts and practical algorithmic implementations. 
\section{Keywords}
  Differential Machine Learning, Risk Neutral valuation, Derivative Pricing, Hilbert Spaces Orthogonal Projection, Generalized Function Theory

\end{abstract}
\newpage

\tableofcontents

\newpage

\section{Introduction}

Within the dynamic landscape of financial modelling, the quest for reliable pricing and hedging mechanisms persists as a pivotal challenge. This article aims to introduce an encompassing theory of pricing valuation uniquely rooted in the domain of machine learning. A primary focus lies in overcoming a prominent hurdle encountered in implementing the differential machine learning algorithm, specifically addressing the critical need for unbiased estimation of differential labels from data sources, as highlighted in studies by Huge (2020) and Broadie (1996). This breakthrough holds considerable importance for contemporary practitioners across diverse institutional settings, offering tangible solutions and charting a course toward refined methodologies. Furthermore, this endeavour not only caters to the immediate requirements of practitioners but also furnishes invaluable insights that can shape forthcoming research endeavours in this domain.

The article sets off from the premise that the pricing and hedging functions can be thought of as elements of a Hilbert space, in a similar way as \cite{pelsser2016difference}. A natural extension of these elements across time, originally attained in the current article, is accomplished by the Hahn Banach extension theorem, an extension that would translate as an improvement of the functional through the means of the incorporation of the accumulating information. This functional analytical approach conveys the necessary level of abstraction to justify, and discuss the different possibilities of implementation of the financial models contemplated in \cite{huge2020differential} and \cite{pelsser2016difference}. So, a bridge will be built from the deepest theoretical considerations into the practicality of the implementations, keeping as a goal mathematical rigour in the exposition of the arguments.
 Modelling in Hilbert spaces allows the problem to be  reduced into two main challenges: the choice of a loss function and the choice of an appropriate basis function. A discussion about the virtues and limitations of two main classes of basis functions is going to unravel, mainly supported by the results in \cite{hornik1989multilayer},\cite{barron1993universal} and \cite{telgarsky2020deep}.
A rigorous mathematical derivation of the loss functions, for the two different risk-neutral methods, is going to be exposed, where the result for the second method,  was stated and proven originally in the current document.
The two methods are the Least Squares Monte Carlo and the Differential machine learning, inspired in \cite{pelsser2016difference} and \cite{huge2020differential}, respectively.
It is noted that the first exposition of the Least Squares Monte Carlo Method was accomplished in \cite{longstaff2001valuing}.
The derivation of the differential machine learning loss function using generalized function theory allows us to relax the assumptions of almost sure differentiability and almost sure Lipschitz continuity of the pay-off function in \cite {broadie1996estimating}. Instead, the unbiased estimate of the derivative labels only requires the assumption of local integrability of the pay-off function, which it must clearly satisfy, given the financial context. This allows the creation of a technique to obtain estimates of the labels for any derivative product, solving the biggest limitation in \cite{huge2020differential}. 
The differential machine learning algorithm efficiently computes differentials as unbiased estimates of ground truth risks, irrespective of the transaction or trading book, and regardless of the stochastic simulation model.

The implementations are going to be completely justified by the arguments developed in the theoretical sections. The implementation of the differential machine learning method relies on \cite{huge2020differential}. The objective of this simulation is to assess the effectiveness of various models in learning the Black-Scholes model within the context of a European option contract. Initially, a comparison will be drawn between the prices and delta weights across various spot prices. Subsequently, the distribution of Profit and Loss (PnL) across different paths will be examined, providing the relative hedging errors metric.
These will serve the purpose of illustrating theoretical developments. 

\section{Set up}
Consider the probability space $(\Omega,\mathscr{F}, Q)$, where $\{\mathscr{F}_t\}^\infty_{t=1}$ is a filtration generated by the adapted stochastic process $\{Z(t,\omega):t\leqslant T\}$, the latter modelling the  underlying asset, where $T\in \mathbb{R}$. $T$ is the expiry date of the derivative on that underlying asset $Z$.

$Q$ is the risk-neutral measure equivalent to the market probability measure $P$, in the Radon-Nykodim sense. Then, $\{Z(t):t\in T\}$ is a local martingale with respect to the measure $Q$.This requires the assumption that $\{Z(t):t\in T\}$ is cádlág uniformly integrable semi-martingale. Applying Doob-Meyer decomposition, and then the Girsanov theorem(see \cite{chung1990introduction} and \cite{beiglboeck2012short}), it is possible to construct a  measure $Q$ such that $\{Z(t,\omega):t\leqslant T\}$
is a local Q-Martingale.
This bears the assumption that a single risk-free short rate is zero.
The paths   $Z(.,w)$ with $\omega\in\Omega$ of $Z$ given by $t\to Z(t,\omega)\:,\: t\in[0,T]$, are assumed to lie in some space $D_d[0,T]$, consisting of functions mapping from $[0,T]$ to $\mathbb{R}^d$. Consider $X$ to be the discounted pay-off function, $\mathscr{F}_T$ measurable. It is clear that the pay-off $X$ depends on the underlying asset $Z$. If the derivative product is path-dependent, for example, an Asian option, the pay-off will depend on the paths from period 0 until period T; otherwise, if it is not path-dependent, the pay-off, denominated as $X$, only depends on $Z(T)$, that is $Z$ at expiry $T$.

A pricing or hedging function space can be defined as $H\subset \mathscr{L}^2(\mathscr{F}_T)$, a complete subspace such that $H_t=H\bigcap \mathscr{L}_2(\mathscr{F}_t)\:,\forall t<T$. The elements in $H$ carry all prior knowledge since it is the intersection of $H$ with $\mathscr{L}^2(\mathscr{F}_T)$, while the elements of $H_t$ carry the knowledge accumulated from time point 0 until time point $t$.It is sensible to assume the restriction that the pricing or hedging function and consequently the pay-off function present themselves with a finite variance, which entails that they are square-integrable, justifying, the previous construction of\:$H$\footnote{H represents all prior knowledge, one of the common constraints for option pricing are non-negativity and positiveness on the second order derivatives}.
\break
A function $g\in H$ defines a pricing or hedging functional acting directly on space of $\mathscr{H}_t$ that can be thought of as a Hilbert space of underlying assets at period $[0,t]$ since it is expected for the asset prices to display a finite variance\footnote{Even when Z is expressed as a diffusion, $T$ is finite so the different paths could never display infinite variance}. The function $g_T$ denotes the pay-off function of a derivative product that expires at time point $T$, while $g_{0,t}\in H_t$ defines the projection of $g_T$ onto $H_t$, for $t<T$.  
\newline
\break
Since the dual of a Hilbert space is itself a Hilbert space,$g$ can be considered a functional.\footnote{This property is easily verified by building the following map $\phi H'\to H$, defined by $\phi(v) = f_v$, where $f_v(x) = \langle x, v\rangle$, for $x \in H$ is an antilinear bijective isometry.}
\break
Considering the sequence of conditional Hilbert spaces:

$$H_t\subset H_l,\forall t<l\leqslant T$$

  the functional $g_{0,t},\in H_t$, can be extended to $H_l=\bigcup_{}^l H_t$, where $g_{0,l}$ denominates its extension, by the Hahn Banach corollaries in \cite{rudin1974functional} and for non-linear version in \cite{wei2020development}.

\hfill \break

Now the pricing or hedging functional incorporates the accumulated information from period $0$ to period $l$.

This allows us to see that the increasing information would shape the function, which is something well-seen, in statistical learning, with the use of increasing training sets defined across time.
\footnote{The functional analytical results can be revisited by the reader in \cite{rudin1974functional}}
\break
We will begin by dwelling upon the problem of how to find function $g$, developing the theoretical statistical objects that are necessary for that aim.
The aim is to estimate the pricing or hedging functions. So, a criterion needs to be established in the theoretical framework.

Let $Z$ and $X$ be two respectively $d$ and $p$ dimensional real-valued random variables, following some unknown joint distribution $p(z,x)$.  The expectation of the loss function associated with a predictor $g$ can be defined as:
\begin{equation}
J(g)\coloneqq\int\int L(g(z),x)p(z,x)
\end{equation}

The objective is to find the element $g\in H$ which achieves the smallest possible expected loss. Assume a certain parameter vector $\theta \in \Theta$, where $\Theta$ is a compact set in the Euclidean space. As the analytical evaluation of the expected value is impossible, a training sample $(z_i,x_i)$ for $i=1,...,n$ drawn from $p(z,x)$ is collected. An approximate solution to the problem can then be found by minimising the empirical approximation of the expected loss:

\begin{equation}
    \min_{\theta\in\Theta}J(\theta)=\min_{\theta\in \Theta}\sum_{i=1}^n L(g_\theta(x_i),y_i)
\end{equation}

\section{From Classical Results into Differential Machine Learning}

\subsection{Risk Neutral Valuation Approach}

The following section discusses two methods: the Least Squares Monte Carlo and the Differential Machine Learning method. The aim is to construct a loss function under the assumption of risk neutrality. We will derive the corresponding loss function and establish two analogous propositions, illustrating the power of the theoretical construction presented earlier. This unique perspective allows for a comprehensive comparison of the two methods, which has not been explored in the existing literature.

As in \cite{cox2000bjork}, the valuation of a derivative contract involves the functional $g_{0,t}$ representing the expected conditional value:

\begin{proposition}
The arbitrage-free price of a claim, whose payoff at time $T$ is $X=g_T(Z(.,\omega))$, is given by:
\begin{equation}
g_{0,t}(Z)=E^{Q}_{t}(X|\mathscr{F}_t)
\end{equation}
where $Q$ is the risk-neutral measure, and $g_{0,t}(Z)$ denotes the expected time $t$ value of $X$.
\end{proposition}

The objective is to compute the expectation of $X$ conditional on information at time $t$. Note that $Q$ is not necessarily unique, meaning there might not be market completeness.

\subsection{Differential Machine learning: building the loss function}

The main interest lies in estimating equation (3) to obtain the price of the derivative. This result contributes to the construction of the Differential Machine Learning approach. The following result is widely known in the literature and can be revisited by the reader in \cite{pelsser2016difference}.

\begin{proposition}
Given $g_{0,t}=E^Q(X|\mathscr{F}_t),\:\forall t\leqslant T$, the loss function is the mean square distance:
\begin{equation}
\int_{\Omega}|X-g_{0,t}|^2d Q
\end{equation}
since
\begin{equation}
E(X|\mathscr{F}_t)=\arg \min_{g_{0,t}\in H_t}\int_{\Omega}|X-g_{0,t}|^2d Q
\end{equation}
\end{proposition}

The proposition that characterizes the Differential Machine learning method is originally stated and proven in the current document. Its proof is essential to understand the implementation of this method, such as how to compute unbiased estimates of the labels for the differentials.

\begin{proposition}
Given $g\in H$ (equation 3) as the risk-neutral valuation formula, and its first-order weak derivative, the loss function regarding an arbitrary time point $t>0$ is:
\begin{equation}
\int_{\Omega}|X-g_{0,t}|^2 dQ+\lambda\int_{\Omega}|\partial X-\partial g_{0,t})|^2 dQ
\end{equation}
since
\begin{equation}
E(X|\mathscr{F}_t)=\arg \min_{g_{0,t}\in H_t}\int_{\Omega}|X-g_{0,t}|^2 dQ+\lambda\int_{\Omega}|\partial X-\partial g_{0,t})|^2 dQ
\end{equation}
where $\partial$ stands for the derivative in the distribution sense and $\lambda>0$ is the Lagrange's constraint.
\end{proposition}

Consider the $g_{0,t}$, the pricing functional as defined in equation (19), that is $g_{0,t}(Z)=E^Q(X|\mathscr{F}_t)$ for some $t<T$, where $T$ is the expiry date.

The $\Delta$ would be computed as:
\begin{equation}
\Delta(z,t)=\frac{\partial E^Q(g_T(Z(T))|\mathscr{F}_t) }{\partial Z(t)}
\end{equation}
which measures the derivative price sensitivity in relation to oscillations of the underlying.

If the differentiation could go inside the conditional expectation, an estimator for $\Delta$ can be created through discretization methods. That would imply that $H$ would need to be restricted to the differentiable elements. An analogous proof to the one of the proposition would be achieved if the space with differentiable elements that have is a separable Hilbert space.

However, pay-off functions are non-differentiable or even discontinuous as in the case of the European standard options and digital options respectively. So a broader concept of derivative needs to be defined. Generalized function theory allows a less abstract approach to weak derivative definition compared to distribution theory as in \cite{rudin1974functional}.

To assure the unbiasedness of the $\Delta$ estimator, there is a need for a result that supports the passage of the derivative inside the expectation. The following result is not a very well-known result and can be found in the seventh chapter of \cite{jones1966generalised} 

\begin{proposition}
Let $X$ be an open set in $\mathbb{R}^n$ and $\Omega$ be a measure space. Given $g(x,\omega)$ for each $\omega\in\Omega$, a generalized function of $x\in X$, define:
\begin{equation}
\langle\int_{\Omega}g(.,\omega)d\omega,\gamma\rangle\coloneqq\int_{\Omega}\langle g(.,w),\gamma\rangle d\omega
\end{equation}
$$\gamma\in \mathscr{D}(x)$$
Assume the above integral is well-defined. Then:
\begin{equation}
\frac{\partial}{\partial x_i}\int_{\Omega}g(x,\omega)dw=\int_{\Omega}\frac{\partial}{\partial x_i}g(x,\omega)d\omega
\end{equation}
where $\frac{\partial}{\partial x_i}$ refers to the generalized derivative of generalized functions on both sides of the equation.
\end{proposition}

\textit{Proof:} Without loss of generality, this will be proven for a two-variable function. Consider $\frac{\partial}{\partial x}=\partial_1$. For an arbitrary good function $\gamma$, $\int_{-\infty}^{\infty}g(x,\mu)\gamma(x) $ defines a generalised function (local integrability), as assumed. Then:
\[
\int_{-\infty}^{\infty}\gamma(x) \bigg\{\partial_1\int_M g(x,w)dw\bigg\}dx=-\int_{-\infty}^{\infty}\partial_1\gamma(x)\ \bigg\{\int_M g(x,w)dw \bigg\}dx\\
\]
\[
=-\int_M\bigg\{\int_{-\infty}^{\infty}g(x,w)\partial_1\gamma(x) dx\bigg\}dw=\int_M\bigg\{\int_{-\infty}^{\infty}\gamma(x)\partial_1 g(x,w) dx\bigg\}dw
\]
proving the assertion.

$\blacksquare$

Assuming that the pay-off function is locally integrable, as it is something expected in financial theory, once the conditional mean of the pay-offs across time is the price of the derivative product, given its idiosyncratic characteristics. This solves the main problem in \cite{broadie1996estimating}, by only requiring the pay-off function to be locally integrable.

Applying proposition 3.4 to equation 8:

$$\Delta(z,x,t)=\frac{ E^Q(\partial g_T(Z)|\mathscr{F_t}) }{\partial Z}$$

The aim now is to construct a space that accounts for the $\partial g(Z,T)$, so that it defines a separable Hilbert space.

This will lead to the theory on the Sobolev spaces, once the latter account for the weak derivatives (derivatives in the distribution sense):

\begin{definition}
So the following space is defined:
\begin{equation}
H^1(\Omega)=\{v\in H,\partial_{z_i}v\in H,\: 1\leqslant i\leqslant d\}
\end{equation}
The space $H^1$ is endowed with the norm associated with the inner product:
\begin{equation}
\langle u,v\rangle=\int_{\Omega}\left(uv + \lambda \sum_{i=1}^d \partial_{z_i}u \partial_{z_i}v\right)dx
\end{equation}
and with the corresponding norm:
\begin{equation}
||u||_{1,\Omega}=\sqrt{\langle u,u \rangle_{1,\omega}}=\int_{\Omega}|u|^2 dx+\lambda\int_{\Omega}|\partial_{z_i} u|^2 dx
\end{equation}
By convention, $H^0=H$.
\end{definition}

It is well known the space above is a Sobolev space and hence it is a complete space as it can be revisited by the reader in \cite{rudin1974functional} and \cite{leoni2017first}.

Now we are in the position to prove Proposition 4.3.

\vspace{0.5cm}
\textit{Proof of Proposition 4.3:}
Consider $H^1$, which accounts for the first-order weak derivative. Considering the pay-off function to be $X=g_{T}$. Since $H^1$ is a Hilbert space, particularly a Sobolev space, applying Proposition 4.4 and the Hilbert Projection Theorem(regarding the later result see \cite{kreyszig1991introductory} ):
\[
E(X|\mathscr{F}_t)=\text{arg min}_{g_{0,t}\in H_t} ||X-g_{0,t}||_{H^1}
\]
\[
=\text{arg min}_{g_{0,t}\in H_t}\int_{\Omega}|X-g_{0,t}|^2 dQ+\lambda \int_{\Omega}|\partial (X-g_{0,t})|^2 dQ
\]
\[
=\text{arg min}_{g_{0,t}\in H_t}\int_{\Omega}|X-g_{0,t}|^2 dQ+\lambda\int_{\Omega}|\partial X-\partial g_{0,t}|^2 dQ
\]
where in the last step, the linearity property of the derivative in the distribution sense was employed.

$\blacksquare$

In equation (7), the second term accounts for the training with respect to $\Delta$, mathematically the shape of the function.

The Hilbert space's separability allows the construction of a countable basis, so $g_{0,t}$ can be expressed as:
$$g_{0,t}=\sum^{\infty}_{k=1}\beta_k v_k$$

Considering the sample: $((x_1,z_1)),...,(x_n,z_n))$. The loss function with respect to the sample is:

\begin{equation}
\begin{split}
L(\theta) = E(X|\mathscr{F}_t) = \lim_{N\to\infty}\lim_{d\to\infty}\text{arg min}_{\theta\in\Theta}\frac{1}{N}\sum_{n=1}^{N}\left(x_n-\sum^{d}_{k=1}\beta_k v_K\right)^2 \\
+ \frac{1}{N}\sum_{n=1}^{N}\left(q_n-\partial\sum^{d}_{k=1}\beta_k v_K\right)^2
\end{split}
\end{equation}

where $q_n$ stands for the differential labels, as will be clearly illustrated in the implementation example.

Before delving into the procedure of building a fixed basis and how it compares with a variable basis, an illustration of an application of Proposition 4.4 is contemplated in the following example.

\section{Example: Digital Options}
\begin{example} In this example, let's verify proposition 4.4 in an application in digital options The aim is to compute $\Delta$.
A digital option is a form of option that allows traders to manually set a strike price. The digital option provides traders with a fixed payout in the case when the market price of the underlying asset exceeds the strike price. 

So the pay-off function can be written as:

$$D(Z(0))=e^{-rT}\mathbb{I}_{Z(T)>K}$$

where $K$ is the strike price, settled upon the celebration of the contract.
Without loss of generality, assume that $r=0$, then:

$$D(Z(0))=\mathbb{I}_{Z(T)>K}$$

Taking the derivative inside the expectation :

$$E(\partial_Z D(Z(0))=E(\partial_Z\mathbb{I}_{Z(T)>K} )=E(-\delta(Z_T-K))$$
where it was used in the last step, that a weak derivative of a heavy side function is the Dirac delta.
Considering the distribution density  to be $f$, whose cumulative distribution is expressed by $F$, then:

$$E(-\delta(Z_T-K))=\int_{-\infty}^{\infty}\delta(z-K)f(z)dz= -f(k)$$

Going the other way around:

$$\partial_Z E(D(Z(0))=\partial_Z E(\mathbb{I}_{Z(T)>K})=$$
$$\partial_Z\int_{K}^{\infty}f(x)dx=\partial_Z\left(\int_{-\infty}^{K}f(z)dz\right)=-f(k)$$

Then, it can be concluded that:

$$E(\partial_Z D(Z(0))=\partial_Z E(D(Z(0))$$
Now, a question could be posed, on how the differential labels can be found. According to this example insofar, we would need to know a priori the density $f$. However, in a scenario where we are presented with data, and not a simulation, an estimator of the expectation of the delta Dirac function on a sample would need to be found. Since delta Dirac would be infinite on a single point, an approximation can be considered. See in \cite{jones1966generalised} that the delta Dirac integral over a test function or a good function is :
\begin{equation}\lim_{n\to\infty}\int_{-\infty}^{\infty}\left(\frac{n}{\pi}\right)^{\frac{1}{2}}e^{-nx^2}\gamma(x)dx=\gamma(0)
\end{equation}

Choosing a large $N$
$$
\int_{-\infty}^{\infty}\left(\frac{N}{\pi}\right)^{\frac{1}{2}}e^{-Nx^2}\gamma(x)dx=\gamma(0)
$$

we finally get an integral we can discretize by the usual numerical methods. This procedure can be applied to any pay-off function, hence to any derivative product, solving the problem in \cite{broadie1996estimating}.
\end{example}

\section{Choice of Basis}

A parametric basis can be thought of as a set of functions made up of linear combinations of relatively few basis functions with a simple structure and depending non-linearly on a set of “inner” parameters e.g., feed-forward neural networks with one hidden layer and linear output activation units. In contrast, classical approximation schemes do not use inner parameters but employ fixed basis functions, and the corresponding approximators exhibit only a linear dependence on the external parameters.

However, experience has shown that optimization of functionals over a variable basis such as feed-forward neural networks often provides surprisingly good suboptimal solutions.

A well-known functional-analytical fact is the employing the Stone-Weierstrass theorem, it is possible to construct several examples of fixed basis, such as the monomial basis, a set that is dense in the space of continuous function whose completion is $\mathscr{L}^2$.
The limitations of the fixed basis are well studied and can be summarized as the following.

\subsection{Limitations of the Fixed-basis}

Assuming a model:
\begin{equation}
    y=f(x)+\epsilon
\end{equation}

where $\epsilon$ is the error term.
\begin{equation}
E(\epsilon^2) = \mathrm{Var}(\hat{f}(x)) + E\left((f(x) - \hat{f}(x))^2\right) + \sigma^2
\end{equation}
The variance-bias trade-off can be translated into two major problems:
\begin{enumerate}
    \item Underfitting happens due to the fact that high bias can cause an algorithm to miss the relevant relations between features and target outputs. This happens with a small number of parameters. In the previous terminology, that corresponds to a low $d$ value (see Equation 4).
    
    \item The variance is an error of sensitivity to small fluctuations in the training set. It is a measure of spread or variations in our predictions. High variance can cause an algorithm to model the random noise in the training data, rather than the intended outputs, which is denominated as overfitting. This, in turn, happens with a high number of parameters. In the previous terminology, that corresponds to a high $d$ value (see Equation 4).
\end{enumerate}

The following result resumes the problem discussed. I will state it as in \cite{barron1993universal}, and the proof can be found in \cite{barron1993universal} and \cite{gnecco2012comparison}
\begin{proposition}

 For every choice of fixed basis functions $h_1,h_2,...,h_n$,
\begin{equation}
\sup_{f\in\gamma_c}d(f,\text{span}(h_1,h_2,...,h_n))\geqslant \kappa\frac{C}{d}\left(\frac{1}{n}\right)^d\end{equation}

where $\kappa$ is a universal constant, not smaller than $1/8\pi e^{\pi-1}$.
Thus, the Kolgomorov n-width of the class of functions $\Gamma_C$ satisfies $$w_n\geqslant \kappa \frac{C}{d}\left(\frac{1}{n}\right)$$

\end{proposition}

where $\Gamma_C$ denotes the class of functions on $\mathbb{R}^d$ that fulfil the following requirements:

$$f(x)=\int_{\mathbb{R}^d}e^{iwx}\tilde{f}(w)dw$$

for some complex valued function $\tilde{f}(w)$, for which $w\tilde{f}(w)$ is integrable. Define:

$$C_f=\int_{\mathbb{R}^d}|w||\tilde{f}(w)|dw$$
\break

where $|w|=\langle w,w\rangle^{1/2}$. For each $C>0$,let $\Gamma_C$ be the set of functions $f$ such that $C_f\leqslant C$.
Since the Fourier transforms on $\mathscr{L}^2(\mathscr{F}_t)$ is an isometry, it can safely be said that $\Gamma_C$ contains $\mathscr{L}_2$ for some arbitrary $C$, the space of interest in the current theoretical development.

Returning to the previous topic of the fixed basis limitations, the distance cannot be made smaller than the error of
order $(1/n)^{2/d}$. As $d$ the number of variables in the target function increases the distance measured,  would increase.  This leads to a well-documented dimensionality curse in \cite{bishop2006pattern}
\newline
\break
So, there is a need to study the class of basis, that can adjust to the data. That is the case with the parametric basis.

\subsection{Parametric Basis: Neural Networks}

In this section, we explore the use of feedforward neural networks as a basis for the function space $H$. We demonstrate that neural networks possess excellent approximation capabilities, analogous to those of the polynomial basis, by leveraging the Stone-Weierstrass theorem. The proofs of the theorems presented in this section can be found in \cite{hornik1989multilayer}.

Firstly, we establish that for any $r\in\mathbb{N}$, the set $\textbf{A}^r$ contains all affine functions from $\mathbb{R}^r$ to $\mathbb{R}$, defined as $A(x)=w\cdot x+b$, where $w$ is a weight vector, $x$ is the input variable vector, and $b\in \mathbb{R}$ is a constant (referred to as bias in neural network terminology). The activation function $G$ plays a crucial role in neural networks, and we define it as follows:

$G:\mathbb{R}\to [0,1]$ is a squashing function if it is non-decreasing, $\lim_{\lambda\to\infty}G(\lambda)=1$, and $\lim_{\lambda\to -\infty}G(\lambda)=0$.

\begin{definition}
For any Borel measurable function $G:\mathbb{R}\to\mathbb{R}$ and $r\in\mathbb{N}$, let $\Sigma\prod^r(G)$ be the class of functions:
\begin{equation}
\{f:\mathbb{R}^r\to \mathbb{R}: f(x)=\sum_{j=1}^{q}\beta_j\prod_{k=1}^{l_j}G(A_{j,k}(x)),\:x\in\mathbb{R}^r,\beta_j\in\mathbb{R},A_{jk}\in \textbf{A}^r\: , q,l_j\in\mathbb{N}\}
\end{equation}
The class of functions in Equation (4) corresponds to the network output function for multi-layer neural networks when $l_j>1$.
\end{definition}

From \cite{hornik1989multilayer}, we find the following relevant results:

\begin{theorem}
    
If $G$ is a continuous non-constant function, then $\Sigma\prod^r(G)$ is uniformly dense on $C(K,\mathbb{R})$, where $K$ is a compact support.
\end{theorem}

\begin{corollary}

If there exists a compact subset $K$ of $\mathbb{R}^r$ such that $\mu(K)=1$, then $\Sigma^r(G)$ is $\rho_p$-dense in $\mathscr{L}^p(\mathbb{R}^r,\mu)$ for every $p\in (1,\infty)$ regardless of $G$, $r$, or $\mu$.
\end{corollary}

From these results, it can be concluded that neural networks are dense in $\mathscr{L}^p$ spaces, including the space of interest $\mathscr{L}^2$. As a result, neural networks constitute a basis for space $H$, with each element expressed as:
\begin{equation}
    v_{i}^{j_t} = A_{Lj} \circ G \circ A_{L-1j} \circ G \circ \ldots \circ A_{1j} \circ G(Z_t)
\end{equation}

The flexibility and approximation power of neural networks makes them an excellent choice as the parametric basis.

\subsubsection{Depth}
In practical applications, it has been noted that a multi-layer neural network, outperforms a single-layer neural network. This is still a question under investigation, once the top-of-the-art mathematical theories cannot account for the multi-layer comparative success.
However, it is possible to create some counter-examples, where the single-layer neural network would not approach the target function as in the following proposition:

\begin{proposition}(Power of depth of Neural Networks \cite{eldan2016power}). There exists a simple
(approximately radial) function on $\mathbb{R}^d$
, expressible by a small 3-layer feedforward neural network,
which cannot be approximated by any 2-layer network, to more than a certain constant accuracy,
unless its width is exponential in the dimension.

\end{proposition}

Therefore it is beneficial or at least risk-averse to select a multi-layer feed-forward neural network, instead of a single-layer feed-forward neural network

\subsubsection{Width}
This section draws inspiration from the works \cite{barron1993universal} and\cite{telgarsky2020deep}. Its primary objective is to investigate the approximating capabilities of a neural network based on the number of nodes or neurons. I provide some elaboration on this result, once it is not so well known and it does not require any assumption regarding the activation function unlike in \cite{barron1994approximation}.

\begin{definition}
An infinite-width shallow network is characterized by a signed measure $\mu$ over weighted vectors in $\mathbb{R}^d$:
\begin{equation}
    x \rightarrow \int \sigma(w^T x) \, du(w)
\end{equation}
\end{definition}

A more general representation, beyond Definition 4.3, can be achieved by considering:
\begin{equation}
    x \rightarrow \int g(x, w) \, d\mu(w)
\end{equation}
where $\mu$ is a signed measure over some abstract parameter space.

We create a probability measure based on $\mu$ as follows:
\begin{enumerate}
    \item By Jordan decomposition, cited from Taylor, $\mu = \mu_+ - \mu_-$ where $\mu_+$ and $\mu_-$ are non-negative measures with disjoint support.
    \item For non-negative measures, we define the total mass $|\mu_{\pm} \|_1 = \mu_{\pm}(\mathbb{R}^p)$, and $\| \mu \|_1 = \| \mu_+ \|_1 + \| \mu_- \|_1$.
\end{enumerate}

Next, we define ${\widetilde{\mu}}$ to sample $s \in \{\pm 1\}$ with $\| 1 \mathop{\textrm{Pr}}[s = +1] = \frac{\|\mu_+\|_1}{\|\mu\|_1} \Pr[s=+1]$, and then sample $\sim \frac{\mu_s}{\|\mu_s\|_1} =: \tilde{\mu}_s$, and output $\tilde{g}(\cdot; w, s) = s \|\mu\|_1 g(\cdot; w)$.

This sampling procedure correctly represents the mean as:
\begin{align}
    \int g(x; w) \, d\mu(w) &= \int g(x; w) \, d\mu_+(w) - \int g(x; w) \, d\mu_-(w) \nonumber \\
    &= \|\mu_+\|_1 \mathbb{E}_{\tilde{\mu}_+} g(x; w) - \|\mu_-\|_1 \mathbb{E}_{\tilde{\mu}_-} g(x; w) \nonumber \\
    &= \|\mu\|_1 \left[ \Pr_{{\widetilde{\mu}}}[s = +1] \mathbb{E}_{{\widetilde{\mu}}_+} g(x; w) - \Pr_{{\widetilde{\mu}}}[s = -1] \mathbb{E}_{{\widetilde{\mu}}_-} g(x; w) \right] \nonumber \\
    &= \mathbb{E}_{\tilde{\mu}} \tilde{g}(x; w, s).
\end{align}

The next proposition is essential in sampling $g$ from $\tilde{\mu}$ and evaluating the distance between the sampled neural network and the true function. A detailed exposition and proof of this proposition can be found in \cite{telgarsky2020deep}.

\begin{proposition}
(Maurey for signed measures).
Let $\mu$ denote a nonzero signed measure supported on $S \subseteq \mathbb{R}^p$, and write $g(x) := \int g(x; w) \, d\mu(w)$. Let $({\tilde{w}}_1, \ldots, {\tilde{w}}_d)$ be i.i.d draws from the corresponding $\widetilde{\mu}$, and let $P$ be a probability measure on $x$. Then:
\begin{align}
    \mathbb{E}_{{\tilde{w}}_1, \ldots, {\tilde{w}}_d} \left\| g - \frac{1}{d} \sum_i \tilde{g}(\cdot; {\tilde{w}}_i) \right\|_{L_2(P)}^2 &\leq \frac{\mathbb{E} \| \tilde{g}(\cdot; {\tilde{w}}) \|_{L_2(P)}^2}{d} \nonumber \\
    &\leq \frac{\|\mu\|_1^2 \sup_{w \in S} \| g(\cdot; w) \|_{L_2(P)}^2}{d},
\end{align}
and moreover, there exist $(w_1, \ldots, w_d)$ in $S$ and $s \in \{\pm 1\}^m$ with
\begin{equation}
    \left\| g - \frac{1}{d} \sum_i \tilde{g}(\cdot; w_i, s_i) \right\|_{L_2(P)}^2 \leq \mathbb{E}_{{\tilde{w}}_1, \ldots, {\tilde{w}}_d} \left\| g - \frac{1}{d} \sum_i \tilde{g}(\cdot; {\tilde{w}}_i) \right\|_{L_2(P)}^2.
\end{equation}
\end{proposition}

As the number of nodes, $d$, increases, the approximation capability improves. This result, contrary to Proposition 5.1, establishes an upper bound that is independent of the dimension of the target function. By comparing both theorems, it can be argued that there is a clear advantage for feed-forward neural networks when $d > 2$ for $d \in \mathbb{N}$.

\section{Simulation-European Call Option}

\subsection{Black-Scholes}

Consider the Black-Scholes model:

\begin{equation}
dS(t)=rS(t)dt+\sigma S(t)dW^Q(t)
\end{equation}

where $S_0=s_0$ and $W$ is a standard Brownian motion as in \cite{black1973pricing}. The parameter values are chosen 
,$S_0 = 100.0$,$K = 110.0$
and $\sigma = 0.2$. The simulation was carried out in Python  The data were simulated according to the risk-neutral diffusion equation(29), by the Euler discretization scheme as in \cite{glasserman2004monte}:

\begin{equation}
    Z^{i}_{T_{j+1}} = Z^{i}_{T_{j}} + r(Z^{i}_{T_{j}},T_{j})(T_{j+1}-T_{j}) + \sigma(Z^{i}_{T_{j}},T_{j})\sqrt{T_{j+1}-T_{j}}N_{j}^{i}
\end{equation}

where i is the index of the path, $j$ is the index of the time step and the $N_j^i$ are independent Gaussian vectors.

\subsection{Hedging Experiment}

This experiment begins by shorting a European call option with maturity $T$. The derivative will be hedged by trading the underlying asset. A $\Delta$ hedging strategy is considered, and the portfolio consisting of positions on the underlying is rebalanced weekly, according to the newly computed $\Delta$ weights.

At maturity, the profit and loss (PnL) are calculated by discrete-time trading on each interval $[t_i, t_{i+1}]$ (where $t_{i+1}-t_i=1/52$ for all $i\leqslant 52$):

\begin{equation}
\text{PnL} = -X + \sum_{i=1}^{52}\Delta_{t_i}(Z_{t_i}-Z_{t_{i-1}})
\end{equation}

A simulation of the evolution of $Z$ across $n$ paths is conducted, producing $n$ PnL values. A histogram is used to visualize the distribution of the PnL values across paths. The different methods are then subjected to this experiment, and the results are compared to the Black-Scholes case.

The PnL values are reported relative to the portfolio value at period 0, which is the premium of the sold European call option. The relative hedging error is measured by the standard deviation of the histogram produced. This metric is widely used in evaluating the performance of different models, as in \cite{frandsen2022delta}.

\subsection{Least Squares Monte Carlo Algorithm}

\subsubsection{Monomial Basis}

As discussed in 5 the monomial basis is a basis of $H_t$ for all $t\in \{0,1,\dots,T\}$. The regression can be written with respect to the monomial basis as:

\begin{equation}
g_T(z_T) = \sum_{i=0}^{d} \beta_i z^i
\end{equation}

where $\beta_i$ is estimated using ordinary least squares.

\subsubsection{Neural Network Basis}

Alternatively, the regression can be conducted using a parametric basis, such as a neural network, where:

\begin{equation}
v_i^j(t,z) = A_{L,j}\circ G \circ A_{L-1,j} \circ G \circ \ldots \circ A_{1,j} \circ G(Z_t)
\end{equation}

and $A_{l,j}(z)=w_{l,j}z+b_{l,j}$ for $l=1,2,\dots,L$. Each element of the basis vector is a function of its inner parameters $w$ and $b$. To find the optimal weights and biases, we minimize the mean square distance between the pay-off and the underlying prices training sample at time point 0:

\begin{equation}
\min_{\theta\in\Theta}J_{LSMC}(\theta) = \min_{\theta\in\Theta}\sum_{j=1}^{n}(x_j-\sum_{i=1}^k b_i v_i(t,z_j|\sigma_j))^2
\end{equation}

The architecture of the neural network is crucial, and a multi-layer neural network with $l>1$ is preferred, as supported by Proposition 5.3 and empirical studies in various applications. In this example, the layer dimension is set to $l=4$, which can be further fine-tuned for explanatory power. The back-propagation algorithm is used to update the weights and biases after each epoch, achieved by minimizing the loss function with respect to the inner parameters through stochastic gradient descent as in \cite{kingma2014adam}
\subsection{Differential Machine Learning Algorithm}

From proposition 3.4, we can infer, given the sample $((x_1,z_1)),\dots,(x_n,z_n))$, that the loss function with respect to the training sample is:
\begin{align}
L(\theta) &= E(X|\mathscr{F}_t) = \lim_{N\to\infty}\lim_{d\to\infty}\text{arg\: min}_{\theta\in\Theta} \frac{1}{N}\sum_{n=1}^{N} \left(x_i-\sum^{d}_{k=1}\beta_k v_K\right)^2 \nonumber\\
&\quad + \frac{1}{N}\sum_{i=1}^{N}\lambda\left( q_i-\sum^{d}_{k=1}\partial \beta_k v_k\right)^2 
\end{align}

The reader might wonder about the meaning of $q_i$. As seen before, the $\{q_i\}_{i}^n$ should be the training sample of the derivative of the pay-off with respect to the underlying at time point $t<T$. To compute $q_i$, we can use equation (8) and Proposition 4.4, which yields:

\begin{equation}
\begin{aligned}
\Delta(z,t) &= \frac{\partial E^Q(g(Z(T))|\mathscr{F}_t) }{\partial Z(t)} \\
&= \frac{\partial E^Q((Z(T)-K)^+)|\mathscr{F}_t) }{\partial Z(t)} \\
&= E^Q\left(\frac{\partial(Z(T)-K)^+)}{\partial Z(t)}|\mathscr{F}_t\right) \\
&= E^Q\left(\frac{\partial(Z(T)-K)^+)}{\partial Z(T)}\frac{\partial Z(T)}{\partial Z(t)}|\mathscr{F}_t\right) \\
&= E^Q(\mathbb{I}_{Z(T)\geqslant K}|\mathscr{F}_t)\frac{\partial Z(T)}{\partial Z(t)}
\end{aligned}
\end{equation}

where the third equality comes from interchanging expectations and differentiation (Proposition 4.4), and the fourth is from applying the chain rule. The (generalized) derivative of the $( )^+$ function is an indicator (or Heaviside) function, so the last factor inside the last expectation above is $\mathbb{I}_{Z(T)}$. Then,
\begin{equation}
q_i=\mathbb{I}_{\{z_i(t)>K\}}\frac{\partial Z(T)}{\partial Z(t)}
\end{equation}

for each individual price at time $t$, across different paths. The local integrability of the pay-off function allows Proposition 4.4 to assure the computation of an unbiased $\Delta$ estimate as above. Simultaneously, the local integrability allows Proposition 4.3 to be applied, rendering the loss function in equation (14). The $\frac{\partial Z(T)}{\partial Z(t)}$ term can be easily calculated in a simulation context and can also be approximated with real data as the rate of increase from the day the contract is celebrated until maturity.

By computing $q_i$ through the simulation of at least two different paths, applying the indicator function, and averaging the resulting quantities, the $\Delta$ hedging estimate is obtained. This method allows for efficient and accurate hedging strategies, making it a valuable tool in the field of mathematical finance.

\subsubsection{Neural Network basis}
All the details of this implementation can be found in \cite{huge2020differential}.
Examining equation (30), the first part is the same loss function as in the LSMC case. Still, the second part constitutes the mean square difference between the differential labels and the derivative of the entire neural network with respect to price. So, we need to obtain the derivative of the feed-forward neural network. Feed-forward neural networks are efficiently differentiated by backpropagation. 

Then recapitulating the feed-forward equations:
\begin{align}
a_0=z_0\\
a_l=W^T G_{L-1}(a_{l-1})+b_l\in\mathbb{R}^{rl},l_j\in\mathbb{N}\\
y=a_L
\end{align}
Recall that the inputs are states and the predictors are prices for the first part, hence, these differentials are predicted risk sensitivities, obtained by differentiation of the line above, in the reverse order:

\begin{align}
 \bar{a}_L=\bar{y}=1  \\
 \bar{a}_l=(\bar{a}_l W^T )\circ G'_{L-1}(a_{l-1})+b_l\in\mathbb{R}^{rl},l_j\in\mathbb{N}
 \\ 
 \bar{a}_0=\bar{z}_0
\end{align}

with the adjoint notation $\bar{x}=\frac{\partial y}{\partial x}$,$\bar{z}_l=\frac{\partial y}{\partial z_l}$,$\bar{y}=\frac{\partial y}{\partial y}=1$

Then the implementation can be divided into the following two steps:
\begin{enumerate}
    \item The neural network for the standard feed-forward equations (35)-(37) is built, paying careful attention to the use of the functionalities of software to store all intermediate values. The neural network architecture will comprehend  4 hidden layers, that is a multi-layer structure as prescribed in section 6.2.1 Note that the activation function needs to be differentiable, in order for equations (38)-(40), to be applied, so the following  \cite{huge2020differential}, a soft-plus function was chosen.

    \item Implement as a standard function in Python the equations (35)-(37). Note that the intermediate values stored before are going to be the domain of this function.

    \item Combine both functions in a single function, named the Twin Tower.

    \item Train the Twin Tower with respect to the loss equation(14).

\end{enumerate}

\section{Numerical Results}

\begin{figure}[H]
  \centering
  \includegraphics[width=0.48\linewidth]{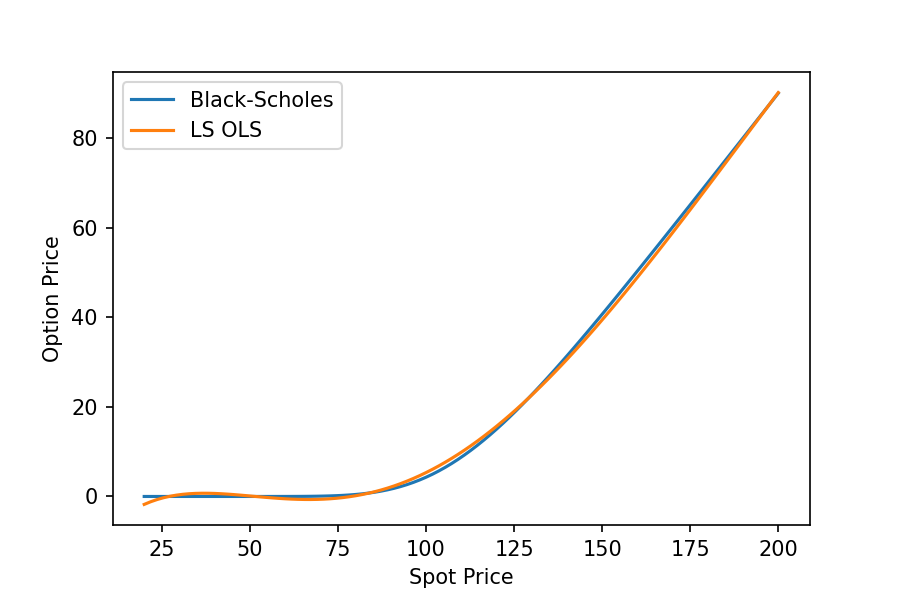}
  \includegraphics[width=0.48\linewidth]{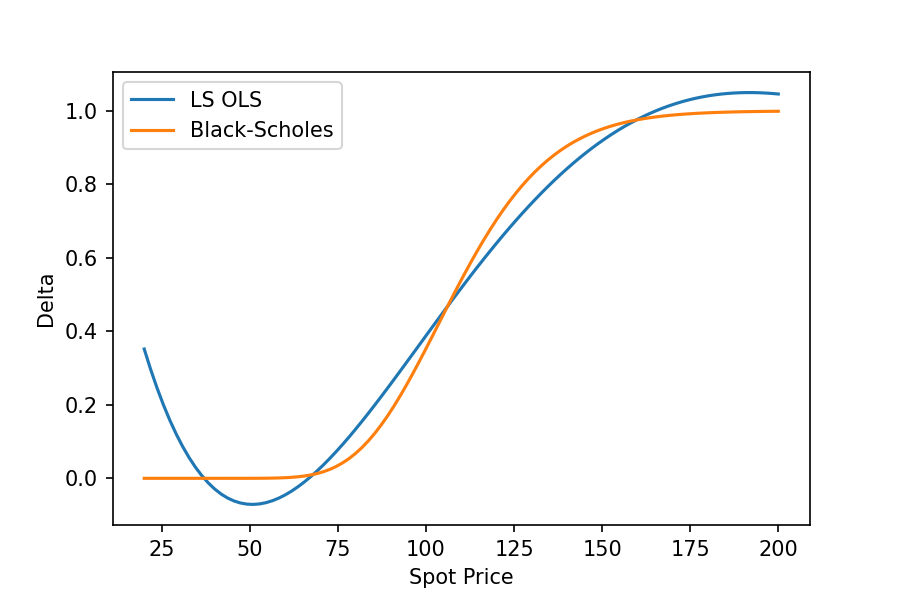}
  \caption{Left: LSMC price estimated with OLS. Right: $\Delta$ hedging function estimated by differentiating the pricing functional (3).}
\end{figure}

\begin{figure}[H]
  \centering
  \includegraphics[width=0.48\linewidth]{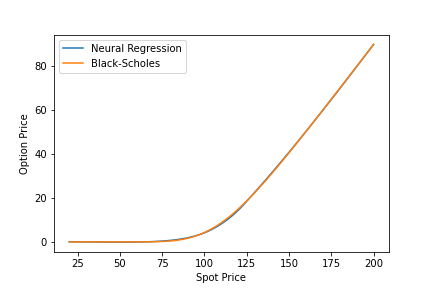}
  \includegraphics[width=0.48\linewidth]{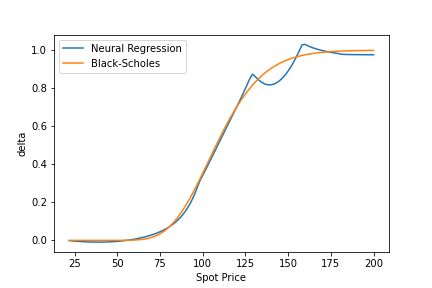}
  \caption{Left: LSMC with multi-layer feed-forward neural network. Right: $\Delta$ hedging from LSMC with multi-layer feed-forward neural network.}
\end{figure}

\begin{figure}[H]
  \centering
  \includegraphics[width=0.48\linewidth]{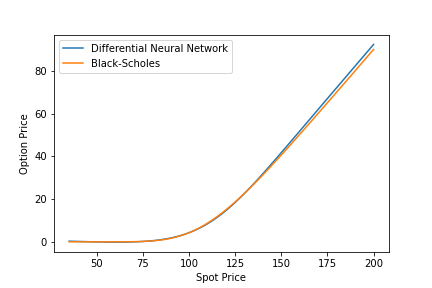}
  \includegraphics[width=0.48\linewidth]{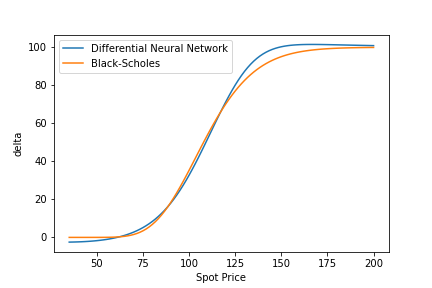}
  \caption{Left: Prices by differential approach with multi-layer feed-forward neural network. Right: $\Delta$ hedging by differential approach with a multi-layer feed-forward neural network.}
\end{figure}

The second part of the simulation analyses will dwell upon the performance of these methods on the hedging experiment, discussed in subsection 6.2.

\begin{figure}[H]
  \centering
  \includegraphics[width=0.48\linewidth]{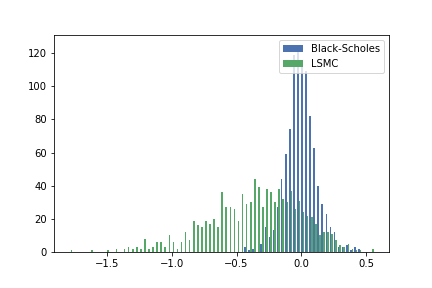}
  \includegraphics[width=0.48\linewidth]{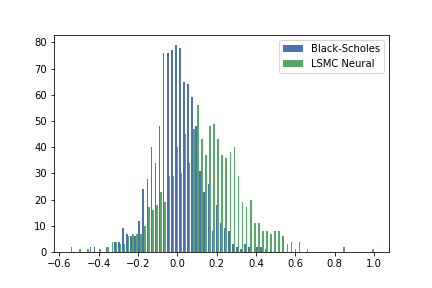}
  \caption{Left: LSMC with polynomial basis PnL distribution against Black Scholes. Right: LSMC with feed-forward neural network basis PnL distribution against Black Scholes.}
\end{figure}

\begin{figure}[H]
  \centering
  \includegraphics[width=0.48\linewidth]{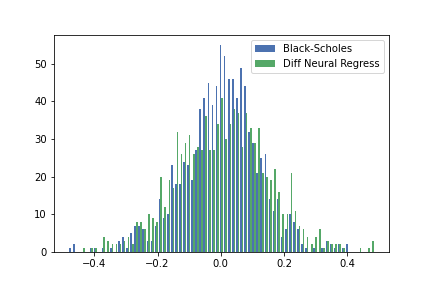}
  \caption{Differential approach with neural networks basis Pnl distribution against Black Scholes.}
\end{figure}

\begin{table}[H]
  \centering
  \begin{tabular}{c c c c c}
    \hline
    \textbf{Training Sample Size} & 1000 & 3000 & 5000 & 7000 \\
    \hline
    Black-Scholes & 0.12 & 0.12 & 0.12 & 0.12 \\
    LSMC Monomial & 0.3340 & 0.2702 & 0.2694 & 0.2694 \\
    LSMC Neural  & 0.20614 & 0.1840 & 0.1725 & 0.1632 \\
    Differential Neural & 0.1449 & 0.1353 & 0.1353 & 0.1353 \\
    
    \hline
  \end{tabular}
  \caption{Relative hedging errors of different methods}
\end{table}

As can be observed in the table, the two implementations at the bottom of the table are the first to converge.

\section{Conclusion}

In conclusion, the strides made in theory have prominently underscored the advantageous application of feed-forward neural networks, with particular emphasis on their integration within the LSMC method. This significance was further validated through the substantial improvements in model fit achieved by employing neural networks as the fundamental framework, as opposed to the explored monomial basis.

Within the domain of risk-neutral models, the standout performer emerges as the differential neural network model. Its augmentation with differentials proved pivotal in seamlessly incorporating shape considerations into the loss function. This distinct advantage translated into superior fits, as highlighted by its remarkable ability to achieve the smallest hedging errors compared to alternative approaches.

The theoretical scaffold constructed here effectively positions the two methodologies – the Least Squares Monte Carlo Method and differential neural networks – as originating from a common premise. However, it's evident that the latter signifies a discernible evolution over the former, delivering a qualitative prediction that harmonizes exceptionally well with the simulation results.

\section{Conflict of Interests Statement}

The author declares that there are no conflicts of interest regarding this research.

\printbibliography

\clearpage

\label{sec:footnotes}
\theendnotes

\listoffigures 

\end{document}